
\input phyzzx

\def\ov{\overline}
\def\un{\underline}

\def\IR{{\hbox{{\rm I}\kern-.2em\hbox{\rm R}}}}
\def\IB{{\hbox{{\rm I}\kern-.2em\hbox{\rm B}}}}
\def\IN{{\hbox{{\rm I}\kern-.2em\hbox{\rm N}}}}
\def\IC{{\ \hbox{{\rm I}\kern-.6em\hbox{\bf C}}}}

\def\IZ{{\hbox{{\rm Z}\kern-.4em\hbox{\rm Z}}}}
\def\to{\rightarrow}
\def\d{{\rm d}}
\def\underarrow#1{\vbox{\ialign{##\crcr$\hfil\displaystyle
{#1}\hfil$\crcr\noalign{\kern1pt
\nointerlineskip}$\longrightarrow$\crcr}}}
%
\def\d{{\rm d}}
\def\ltorder{\mathrel{\raise.3ex\hbox{$<$}\mkern-14mu
             \lower0.6ex\hbox{$\sim$}}}
\def\lesssim{\mathrel{\raise.3ex\hbox{$<$}\mkern-14mu
             \lower0.6ex\hbox{$\sim$}}}


\def\overlrarrow#1{\vbox{\ialign{##\crcr
$\leftrightarrow$\crcr\noalign{\kern-1pt\nointerlineskip}
$\hfil\displaystyle{#1}\hfil$\crcr}}}
\def\un{\underline}
\def\ov{\overline}
\overfullrule=0pt
\tolerance=5000
\overfullrule=0pt
\twelvepoint

\twelvepoint
\pubnum{IASSNS-HEP-93/29}
\date{June, 1993}
\titlepage
\title{QUANTUM BACKGROUND INDEPENDENCE \break
\break IN STRING THEORY}
\vglue-.25in
\author{Edward Witten
\foot{Research supported in part by NSF Grant
PHY92-45317.}}
\medskip
\address{School of Natural Sciences
\break Institute for Advanced Study
\break Olden Lane
\break Princeton, NJ 08540}
\bigskip
\abstract{Not only in physical string theories, but also in some highly
simplified situations, background independence has been
difficult to understand.  It is argued that the ``holomorphic
anomaly'' of Bershadsky, Cecotti, Ooguri, and Vafa gives a fundamental
explanation of some of the problems.
Moreover, their anomaly equation can be interpreted
in terms of a rather peculiar quantum version of background independence:
in systems afflicted by the anomaly,
background independence does not hold order by order in
perturbation theory, but the exact partition function as a function of the
coupling constants has a background independent interpretation as
a state in an auxiliary quantum Hilbert space.  The significance of this
auxiliary space is otherwise unknown.}
\endpage

\chapter{Background Independence And The Holomorphic Anomaly}

Finding the right framework for an
intrinsic, background independent formulation of string theory is
one of the main problems in the subject, and so far has remained out
of reach.  Moreover, some highly simplified special
cases or analogs of the problem, which look like they might be studied
for practice, have also resisted understanding.

An important example is the problem of understanding the mirror map in
the theory of mirror symmetry.
In $(2,2)$ compactification on a Calabi-Yau threefold $X$, one encounters
two sets of renormalizable Yukawa couplings, involving modes
coming from $H^{1,1}(X)$ and $H^{2,1}(X)$.  These are closely related
to two twisted topological field theories that can be constructed for
a given Calabi-Yau target space $X$ -- the $A$ model and the $B$ model.

Mirror symmetry is a relation
between two Calabi-Yau manifolds $X$ and $Y$, in which the Yukawa
couplings involving $H^{1,1}(X)$ are identified with those that
involve $H^{2,1}(Y)$, and vice-versa.  Equivalently, mirror symmetry
exchanges the $A$ model of $X$ with the $B$ model of $Y$, and vice-versa.

The moduli space of sigma models with a Calabi-Yau target space is locally
a product of two factors.  One factor, the moduli space ${\cal M}_A$ of
the $A$ model,  is (an open set in)
$H^{1,1}(X,\IC/2\pi i \IZ)$.  The other factor, the moduli space ${\cal M}_B$
of
the $B$ model, is the moduli space of complex structures on $X$.
Mirror symmetry therefore implies a natural map between ${\cal M}_A(X)$ and
${\cal M}_B(Y)$.  This seems bizarre because -- being related to the linear
space $H^{1,1}(X,\IC)$ --
${\cal M}_A$ has a natural
``flat'' (or really, affine linear) structure,
while ${\cal M}_B$, the moduli space of complex structures, has no
such natural structure.

\REF\candelas{P. Candelas, X.C. de la Ossa,       P. S. Green, and
L. Parkes, ``A Pair Of Calabi-Yau Manifolds As An Exactly Soluble
Superconformal Theory,'' Nucl. Phys. {\bf B359} (1991) 21.}
\REF\griffiths{R. Bryant and P. Griffiths, in {\it Arithmetic and Geometry},
vol. II. p. 77 (M. Artin and J. Tate, eds.), Birkhauser, Boston (1983).}
\REF\morrison{D. Morrison, ``Mirror Symmetry And Rational Curves On
Quintic Threefolds: A Guide For Mathematicians,'' J. Am.
Math. Soc. {\bf 6} (1993) 223.}
\REF\agm{P. Aspinwall, B. Greene, and P. Morrison, ``Multiple Mirror
Manifolds And Topology Change In String Theory,'' IASSNS-HEP-93/4.}
Candelas et. al. [\candelas] tried to overcome this problem by using
the ``special coordinates'' on ${\cal M}_B$.  Special coordinates are
background dependent.  They are defined as follows.
Pick a complex structure $J_0$ on $X$ representing a
base-point in ${\cal M}_B$.  Let $\Omega_0$ be a
three-form on $X$ holomorphic with respect to the complex structure $J_0$.
Let $J$ be a variable complex structure on $X$, and let $\Omega_J$ be
a holomorphic three-form in the complex structure $J$, normalized
so that if one decomposes $\Omega_J$ with respect to the Hodge structure
defined by $J_0$, then the $(3,0)$ part of $\Omega_J$ is cohomologous to
$\Omega_0$.  Then [\griffiths] the map from ${\cal M}_B$ to the $(2,1)$
part of the cohomology class of
$\Omega_J$ is locally an isomorphism from ${\cal M}_B$ to the
linear space $H^{2,1}(X_0,\IC)$ (here $X_0$ is $X$ with complex structure
$J_0$).
\foot{Upon contraction with $\Omega_0$, $H^{2,1}(X_0,\IC)$ can be identified
with $H^1(X_0,T^{1,0}X_0)$,
which is the tangent space to ${\cal M}_B$ at its chosen base-point $X_0$.
This is the natural linear space with which one might try to identify
${\cal M}_B$ near $X_0$.}
This gives the desired ``flat'' structure on ${\cal M}_B$, the
``special coordinates'' being the components of the $(2,1)$ part of
$\Omega_J$.

The trouble with this is of course the dependence on the base-point $J_0$.
According to Morrison [\morrison], the mirror of the natural flat
structure on ${\cal M}_A(X)$ is the flat structure determined on ${\cal
M}_B(Y)$
by a base-point at infinity.  In this context, ``infinity'' refers to
a particular type of degeneration of the complex structure of $X$; neither
existence nor uniqueness of such a degeneration is apparent.
(Lack of uniqueness can lead to a multiple mirror phenomenon and
topology change [\agm].)  So we come to our first problem:

(1) What is the analog in the $A$ model of choosing a base-point in the
$B$ model?  Why does the obvious flat structure on the parameter space of
the $A$ model correspond to the $B$ model with a base-point at infinity?

\REF\witten{E. Witten, ``Chern-Simons Gauge Theory As A String Theory,''
IASSNS-HEP-92/45, to appear in the Floer Memorial Volume.}
Apart from using Calabi-Yau threefolds for compactification of physical
string theories, one can use the twisted $A$ or $B$ topological field
theories to construct topological string theories which one might
think would be a highly simplified laboratory for studying background
independence in string theory.  Indeed, for open string versions of either
the $A$ or $B$ model,
there is no problem [\witten] in identifying the background independent
space-time physics, which moreover is local in some important cases.
But the natural attempt at extracting an effective space-time field theory
for closed topological strings
(see \S5 of [\witten]) gives a result that is non-local and is background
dependent in the case of the $B$ model.  So, even if we put aside the
non-locality, we have our second problem of background independence:

(2) What is the origin of the background dependence in the space-time
field theory of the closed string $B$ model?

A special case of these problems is so simple that it seems worthy of
pointing out separately.  The usual analysis of the $B$ model appears
to show that in genus zero, in expanding around an arbitrary base-point $J_0$,
the partition function vanishes together with its first two derivatives
(while the third derivative gives the Yukawa couplings).
So our third question is obviously:

(3) Since background independence appears to require that the choice of
base-point $J_0$ should play no special role, does not vanishing of the
genus zero partition function at $J_0$
require that function to vanish identically?

\REF\li{K. Li, ``Topological Gravity With Minimal Matter,'' Nucl. Phys.
{\bf B354} (1991) 711, ``Recursion
Relations In Topological Gravity With Minimal Matter,'' Nucl. Phys.
{\bf B354} (1991) 725.}
\REF\warner{M. Bershadsky, W. Lerche, D. Nemeschansky, and N. P. Warner,
``Extended $N=2$ Superconformal Structure Of Gravity and $W$-Gravity
Coupled to Matter,'' CALT-68-1832.}
\REF\mukhi{S. Mukhi and C. Vafa, ``Two Dimensional Black Hole As A
Topological Coset Of $c=1$ String Theory,'' HUTP-93/A002.}
\REF\panda{S. Panda and S. Roy, ``On The Twisted $N=2$ Superconformal
Algebra In 2D Gravity Coupled To Matter,'' IC/93/81.}
\REF\vafa{M. Bershadsky, S. Cecotti, H. Ooguri, and C. Vafa,
``Holomorphic Anomalies In Topological Field Theories,''
HUTP-93/A008,RIMS-915.}
\REF\cecotti{S. Cecotti, P. Fendley, K. Intrilligator, and C. Vafa,
``A New Supersymmetric Index,'' Nucl. Phys. {\bf B386} (1992) 405.}
\REF\cecottti{S. Cecotti and C. Vafa, ``Ising Model And N=2 Supersymmetric
Theories,'' HUTP-92/A044, SISSA-167/92/EP (1992).}
Another cluster of questions that may be similar involves soluble
string theories in $D\leq 2$.  At least in the case of $D=2$, where
one has a two dimensional space-time, a graviton-dilaton system
with a black hole solution, a tachyon scattering matrix, etc., one would
hope to find an intrinsic description of the space-time geometry of these
theories.  Yet this has proved surprisingly elusive.
A possible relation of these problems to those discussed above is suggested
by the fact that the $D\leq 2$ string theories can be interpreted
via twisted $N=2$ models, as first shown for $D<2$ by K. Li [\li] and
argued more recently for $D=2$ [\warner--\panda].

So our final problem -- which we will actually not discuss in this
paper -- is the following:

(4) Either describe the background independent space-time physics
of $D=2$ (or maybe $D\leq 2$) string theory or describe the obstruction
to doing so.

\section{Role Of The Holomorphic Anomaly}

Recently, Bershadsky, Cecotti, Ooguri, and Vafa [\vafa], following earlier
work [\cecotti,\cecottti], described
a ``holomorphic anomaly'' in topological field theories obtained
by twisting $N=2$ models.  Their anomaly can be understood
as a violation of naive background independence and
I will presently argue that it explains most of the puzzles cited above.

\def\bar{\overline}
Consider an $N=2$ supersymmetric theory that contains chiral or twisted
interactions, in addition to
other interactions that we will bury in a Lagrangian $L_0$.
There is no essential
loss in considering the chiral case.  If $W_a$ are the chiral primary
fields, then the Lagrangian
$$L = L_0-\sum_a t^a\int \d^2x\,\d^2\theta \,W_a
-\sum_a\overline t^a \int \d^2x\,\d^2\overline\theta\,\bar W_a,\eqn\ucu$$
with complex parameters $t^a$, describes a family of $N=2$ theories.
One can also consider twisted topological field theories with
two of the supersymmetries, the ones that generate shifts in $\bar\theta$,
identified with BRST operators $Q_\pm$.
Then the $W_a$ are the physical observables, and
$\int\d^2x\,\d^2\bar\theta\, \bar W_a$ is formally irrelevant being
$\{Q_+,[Q_-,\ov W_a]\}$.  After twisting \ucu, it is natural to add the
physical observables as perturbations, considering a more general
Lagrangian
$$L=L_0-\sum_a (t^a+u^a)\int\d^2x\,\d^2\theta \, W_a
-\sum_a\bar t^a\int\d^2x\d^2\bar\theta \,\,\,\bar W_a.\eqn\hucu$$

Formally, the BRST machinery appears to show that the topological
observables of the topological field theory \hucu\ are independent
of $\bar t^a$, and so are only functions of $t^a+u^a$.  This
is where the holomorphic anomaly comes in; Bershadsky et. al.
show that the topological observables really have a dependence on
$\bar t^a$ determined by their holomorphic anomaly.  I want to interpret
this as a failure of background independence.
The idea is that the choice of $\bar t^a$ determines the original
``physical'' Lagrangian \ucu,
which was then twisted and perturbed by topological observables with
coefficients $u^a$.  The dependence on $\bar t^a$ means that the particular
``physical'' theory that one started with is not forgotten.
It defines a base-point in the space of theories.

Let us change the notation, renaming $t+u$ as $t$ and $\overline t$
as $t'$.  Henceforth $\bar t$ denotes the complex conjugate of $t$; $t'$
is an independent complex variable.  Now \hucu\ takes a more
symmetric-looking form,
$$L=L_0-\sum_a t^a \int\d^2x\,\d^2\theta \, W_a
-\sum_a t'{}^a\int\d^2x\d^2\bar\theta \,\,\bar W_a.\eqn\hucu$$
The symmetry between $t$ and $t'$ is broken by the choice of twisting.
The $t'{}^a$ determine a base-point in the space of topological
couplings, the point at which $\bar t^a=t'{}^a$ and at which the theory
can actually be interpreted as a twisting of a physical theory.
The BRST argument appears to show that in the twisted model, the
topological observables depend
only on the $t^a$ and not on the $t'{}^a{}$, but the anomaly obstructs
this.

Now we can rather easily dispose of the first three problems identified
above:

(1) The symmetry in structure between $A$ and $B$ models is restored
because just as the $B$ model naturally depends on a base-point
in the space of complex structures, the $A$ model also
depends on a choice of base-point, namely a choice of a point $t'{}^a
\in H^{1,1}(X,\IC/2\pi i \IZ)$.   There is therefore a more general
family of $A$-like models than hitherto realized.  Bershadsky et. al.
show in detail that the traditional $A$ model (in which correlation functions
are given by standard instanton sums) corresponds to the case in which
$t'{}^a\to \infty$, that is the case in which the $A$ model is taken
to have a base-point at infinity.  The mirror
of the traditional $A$ model is therefore naturally
a $B$ model with base-point at infinity.
A more general $B$ model with another base-point would be mirror to an
$A$ model with a finite base-point $t'{}^a$.

(2) Because of the holomorphic anomaly,
background independence of the field theory of the $B$ model is not
expected.

(3) The paradox involving the genus zero free energy $F_0$ is
also eliminated once one abandons naive background independence.
One is merely left with the statement that if the model is constructed
with a base-point $t'{}^a$, then $F_0$ vanishes together with its
first two derivatives near $t^a=\bar t'{}^a$.

The fourth problem on our above list -- the space-time physics
of soluble string theories -- should perhaps also be reexamined in
light of the holomorphic anomaly, and its possible cousins.

\section{Salvaging Something From Background Independence}

Though the interpretation of the holomorphic anomaly as an obstruction
to background independence eliminates some thorny puzzles,
it is not satisfactory to simply leave matters at this.
Is there some more sophisticated
sense in which background independence does hold?

In thinking about this question, it is natural to examine
the all orders generalization
of the holomorphic anomaly equation, which (in the final equation of their
paper) Bershadsky et. al. write in the following form.  Let $F_g$ be
the genus $g$ free energy.  Then
$$\bar\partial_{i'}
F_g=\bar C_{i'j'k'}e^{2K}G^{jj'}G^{kk'}\left(
 D_{j}D_kF_{g-1}+
{1\over 2}\sum_rD_j F_r  \,\cdot\,D_kF_{g-r}\right).\eqn\murgo$$
This equation can be written as a linear equation for
$Z=\exp\left({1\over 2}\sum_{g=0}^\infty \lambda^{2g-2} F_g\right)$,
namely
$$\left(\bar\partial_{i'}-\lambda^2\bar C_{i'j'k'}e^{2K}G^{jj'}G^{kk'}
D_jD_k\right) Z = 0 . \eqn\hiko$$
This linear equation is called a master equation by
Bershadsky et. al.; it is similar in structure to the heat equation obeyed
by theta functions.  (I am here using the notation of Bershadsky et. al.,
but later, we will make some changes in notation.)

It would be nice to interpret \hiko\ as a statement of some sophisticated
version of background independence.  In thinking about this question,
a natural analogy arises with Chern-Simons gauge theory in $2+1$ dimensions.
In that theory, an initial value surface is a Riemann surface $\Sigma$.
In the Hamiltonian formulation of the theory, one constructs a Hilbert
space ${\cal H}$ upon
quantization on $\Sigma$.  ${\cal H}$ should be obtained by quantizing
a certain classical phase space ${\cal W}$ (a moduli space of flat
connections on $\Sigma$). Because the underlying Chern-Simons
Lagrangian does not depend  on a choice of metric,
one would like to construct ${\cal H}$ in a natural, background independent
way.  In practice, however, quantization of ${\cal W}$ requires a choice
of polarization, and there is no natural or background independent
choice of polarization.

The best that one can do is to pick a complex structure $J$ on $\Sigma$,
whereupon ${\cal W}$ gets a complex structure.
Then a Hilbert space
${\cal H}_J$ is constructed as a suitable space of holomorphic functions
(really, sections
of a line bundle) over ${\cal W}$.  We denote such a function as
$\psi(a^i;t'{}^a)$ where $a^i$ are complex coordinates on ${\cal W}$ and
$t'{}^a$ are coordinates parametrizing the choice of $J$.

\REF\seiberg{S. Elitzur, G. Moore, A. Schwimmer, and N. Seiberg,
``Remarks On The Canonical Quantization Of The Chern-Simons-Witten Theory,''
Nucl. Phys. {\bf B326} (1989) 108.}
\REF\axelrod{S. Axelrod, S. DellaPietra, and E. Witten, ``Geometric
Quantization Of Chern-Simons Gauge Theory,'' J. Diff. Geom. {\bf 33} (1991)
787.}
Now background independence does not hold in  a naive sense; $\psi$ cannot be
independent of $t'{}^a$ (given that it is to be holomorphic on ${\cal W}$ in a
complex structure that depends on $t'{}^a$).  But there is a more
sophisticated sense in which background independence can be formulated
[\seiberg,\axelrod].
The ${\cal H}_J$'s can be identified with each
other (projectively) using a (projectively) flat connection over the space of
space of $J$'s.  This  connection $\nabla$ is such that a covariantly constant
wave function should have the following property:
as $J$ changes, $\psi$ should
change by a Bogoliubov transformation, representing the effects of a change
in the representation used for the canonical commutation relations.

Using parallel transport by $\nabla$ to identify the
various ${\cal H}_J$'s, one can speak of ``the'' quantum Hilbert
space, of which the ${\cal H}_J$'s are realizations determined by a
$J$-dependent choice of representation of the canonical commutators.
Background independence of $\psi(a^i;t'{}^a)$ should be
interpreted to mean that the quantum state represented by $\psi$ is
independent of the $t'{}^a$, or equivalently that $\psi $ is invariant under
parallel transport by $\nabla$.
Concretely, this can be written as an equation
$$0=\left({\partial\over\partial t'{}^a}-{1\over 4}\left(
{\partial J\over\partial t'{}^a}{\omega^{-1}}
\right)^{ij}{D\over Da^i}{D\over D a^j}
\right)\psi            \eqn\mucuu$$
that is analogous to the heat equation for theta functions.
We will discuss this equation further in \S2.

\hiko\ and \mucuu\ have an obvious similarity.  Our goal in the rest
of this paper will be to make this similarity more precise, introducing
an auxiliary quantum system with quantum Hilbert space ${\cal H}$
and interpreting \hiko\ as the statement that the vector in ${\cal H}$
determined by the partition function $Z$ is independent of a choice of
polarization.  For $X$ a Calabi-Yau threefold, the phase space of
the auxiliary system will be simply ${\cal W}=H^3(X,\IR)$.  ${\cal W}$
has a natural symplectic structure given by the intersection pairing
$$\omega(\alpha,\beta)=\int_X\alpha\wedge \beta. \eqn\jcjc$$
${\cal W}$ has no natural complex structure, but every choice of
complex structure on $X$ determines a complex structure on ${\cal W}$
via the Hodge decomposition.  Then it turns out (at least up to a $c$-number
factor) that \hiko\ can be
interpreted to mean that the quantum state determined by the partition
function $Z$ is independent of the base-point.

Though this interpretation of the holomorphic anomaly is elegant,
its rationale remains obscure.  What really is the origin of the phase
space ${\cal W}$, what is the significance of the Hilbert space ${\cal H}$,
and why should it be possible to interpret the partition function as
a $J$-independent vector in ${\cal H}$?  In the case of Chern-Simons
theory, these questions are answered by appealing to the underlying
three dimensional Chern-Simons action and field theory, but in the present
case,
it is not clear where an answer would come from.

\chapter{Quantum Background Independence}

\section{Quantization}

Before plunging into our specific problem, I will first recall
some generalities about quantization of linear spaces.
(See the introduction to [\axelrod] for more detail.)
We consider
a linear space ${\cal W}\cong
\IR^{2n}$ with a constant symplectic structure, say $\omega={1\over 2}
\omega_{ij}\d x^i\d x^j$, with $\omega_{ij}$ a constant invertible
matrix and the $x^i$ linear coordinates on $\IR^{2n}$.
The symbol $\omega^{-1}$ will denote the matrix inverse to $\omega$,
obeying $\omega_{ij}\omega^{-1}{}^{jk}=\delta_i{}^k$.
\REF\kostant{B. Kostant, {\it Orbits, Symplectic Geometry, and Representation
Theory}, Proc. U.S.-Japan Seminar on Differential Geometry (Kyoto, 1965),
{\it Quantization And Unitary Representations}, Lecture Notes In Mathematics,
Vol. 170 (Springer, Berlin, 1970), ``Line Bundles And The Prequantized
Schrodinger Equation,'' Coll. Group Theoretical Methods in Physics,
Marseille (1972) p. 81.}
\REF\souriau{J.-M. Souriau, {\it Quantification Geometrique}, Comm. Math.
Phys. {\bf 1} (1966) 374, {\it Structures Des Systems Dynamiques}
(Dunod, Paris, 1970).}
\REF\woodhouse{N. Woodhouse, {\it Geometric Quantization} (Oxford Univ. Press,
Oxford, 1980).}
\REF\sniatycki{J. Sniatycki, {\it Geometric Quantization And Quantum
Mechanics} (Springer, New York, 1980).}
To give the problem of quantization its most natural (but perhaps
not entirely familiar) formulation [\kostant--\sniatycki],
one begins by introducing a ``prequantum line bundle''; this is a unitary
line bundle ${\cal L}$ with a connection whose curvature is $-i\omega$.
Up to isomorphism, there is only one such choice of ${\cal L}$.   One can
take ${\cal L}$ to be the trivial unitary line bundle, with a connection
given by the covariant derivatives
$$ {D\over Dx^i}= {\partial\over\partial x^i}+{i\over 2}\omega_{ij}x^j.
\eqn\hsoh$$
Then one can introduce the ``prequantum Hilbert space'' ${\cal H}_0$
which consists of ${\bf L}^2$ sections of ${\cal L}$.

A vector in ${\cal H}_0$ is represented by a function with a rather
general dependence on all $2n $ coordinates $x^i$.  The quantum Hilbert
space is instead to be a comparatively ``small'' subspace of ${\cal H}_0$
consisting of functions that depend freely on only $n$ of the coordinates.
There is no natural way to choose which $n$ coordinates are allowed;
such a choice is called a choice of polarization.

We will consider a polarization defined by a choice of a complex structure
$J$ on ${\cal W}$ with the following properties:

(a) $J$ is translation invariant, so it is defined by a constant
matrix $J^i{}_j$ with $J^2=-1$.

(b) The two-form $\omega$ is of type $(1,1)$ in the complex structure
$J$. In components this means that $J^{i'}{}_{i}J^{j'}{}_j\omega_{i'j'}=\omega
_{ij}$ or equivalently (using $J^2=-1$)
$$ J^{i'}{}_i\omega_{i'j} = -J^{j'}{}_j\omega_{ij'}. \eqn\mcon$$
Since the curvature of the prequantum line bundle ${\cal L}$ is proportional
to $\omega$, having $\omega$ be of type $(1,1)$ means that the $(0,2)$
part of the curvature vanishes,
so that the connection on ${\cal L}$ endows it with a complex structure.

(c) $J$ is positive in the sense that the metric $g$ defined by
$g(v,w)=\omega(v,Jw)$ is positive.

Given such a $J$, we define the quantum Hilbert space ${\cal H}_J$
to consist of vectors in ${\cal H}_0$ that are represented by functions
that are holomorphic in the complex structure $J$.

${\cal H}_J$ is a quantization of the symplectic manifold ${\cal W}$;
we want to exhibit a flat connection over the parameter space of the $J$'s
that will enable us
to identify the
${\cal H}_J$'s.  Construction of such a connection enables one to
speak of ``the'' quantum Hilbert space ${\cal H}$ which has realizations
depending on the choice of $J$.  Actually, the connection will only be
projectively flat, so this will only work up to a scalar multiple.

\def\un{\underline}
\def\ov{\overline}
To write down the connection, some notation is useful.  First of all,
one has projection operators
$${1\over 2}\left(1\mp i J\right) \eqn\obo$$
that project onto the $(1,0)$ and $(0,1)$ parts of a vector.  It
is convenient (as in [\axelrod]) to introduce a special notation for
vectors that have been so projected.  For any vector $v^i$, we write
$$\eqalign{ v^{\un i}= & {1\over 2}\left(1-iJ\right)^i{}_jv^j\cr
            v^{\ov i}= & {1\over 2}\left(1+iJ\right)^i{}_jv^j.\cr}
               \eqn\mango$$
Similarly for a one-form $w_i$, we write
$$\eqalign{ w_{\un j}= & {1\over 2}\left(1-iJ\right)^i{}_jw_i\cr
            w_{\ov j}= & {1\over 2}\left(1+iJ\right)^i{}_j w_i.\cr}
               \eqn\mango$$
For example, $J$ itself has non-zero components $J^{\un i}{}_{\un j}
=i\delta^{\un i}{}_{\un j}$,  $J^{\ov i}{}_{\ov j}=-i\delta^{\ov i}{}_{\ov j}
$ (which means that projections of $J^i{}_j$ and $\delta^i{}_j$
are proportional).

Let ${\cal M}$ be the
space of $J$'s obeying conditions (a), (b), and (c) above; it is a copy of
the Siegel upper half plane.  ${\cal M}$ has a natural complex structure,
defined as follows.  The condition $J^2=-1$ implies that for $\delta J$ a
first order variation of $J$, one must have $J\cdot \delta J+\delta J\cdot J
=0$.  This means that the non-zero projections of $\delta J$ are
$\delta J^{\un i}{}_{\ov j}$ and $\delta J^{\ov i}{}_{\un j}$.  We give
${\cal M}$ a complex structure by declaring
$\delta J^{\un i}\,{}_{\ov j}$ to be of type $(1,0)$
and $\delta J^{\ov i}{}_{\un j}$ to be of type $(0,1)$.

Over ${\cal M}$ we now introduce two Hilbert space bundles.
One of them, say ${\cal H}^0$, is the trivial bundle ${\cal M}
\times {\cal H}_0$ whose fiber is the fixed Hilbert space ${\cal H}_0$.
(Recall that the definition of ${\cal H}_0$ was independent of $J$.)
The second is the bundle, say ${\cal H}^Q$, whose fiber over a point
$J\in {\cal M}$ is the Hilbert space ${\cal H}_J$.  ${\cal H}^Q$ is
a sub-bundle of ${\cal H}^0$; a section of ${\cal H}^0$ is an arbitrary
function  $\psi(x^i;J)$, while a section of ${\cal H}^Q$ is a
$\psi(x^i;J)$ which for each given $J$ is, as a function of the $x^i$,
holomorphic in the complex structure defined by $J$:
$${D\over D x^{\ov i}} \psi = 0 . \eqn\jsnsn$$
(This equation has a dependence on $J$ coming from the projection
operators used in defining $x^{\ov i}$.)

A connection on ${\cal H}^0$ restricts to a connection on ${\cal H}^Q$ if
and only if its commutator  with $D_{\ov i}$ is a linear combination of
the $D_{\ov j}$.  For instance, since
${\cal H}^0$ is defined as the product bundle ${\cal M}\times {\cal H}_0$,
there is a trivial connection $\delta$ on this bundle:
$$\delta =\sum_{i,j}\d J^i{}_j{\partial\over\partial J^i{}_j}.\eqn\xnx$$
Thus, $\psi(x^i;J)$ is annihilated by $\delta $ if and only if it is
independent of $J$.  We can expand $\delta$ in $(1,0)$ and $(0,1)$ pieces,
$\delta=\delta^{(1,0)}+\delta^{(0,1)}$, with
$$\eqalign{ \delta^{(1,0)} & = \sum_{i,j}\d J^{\un i}\,{}_{\ov j}{\partial
\over \partial J^{\un i}{}_{\ov j}} \cr
\delta^{(0,1)} & = \sum_{i,j}\d J^{\ov i}{}_{\un j}{\partial
\over \partial J^{\ov i}{}_{\un j}}. \cr} \eqn\concon$$
The commutation relation
$$\left[\delta, D_{\ov i}\right] =\left[\delta,{1\over 2}\left(D_i
+iJ^k{}_iD_k\right)\right]={i\over 2}\d J^k{}_iD_k \eqn\oncon$$
shows that $[\delta, D_{\ov i}]$ is not a linear combination of the
$D_{\ov j}$ and hence that $\delta$ does not descend to a connection
on ${\cal H}^Q$.

Rather, we must add an extra term that reflects the effects of the
Bogoliubov transformation on the quantum state.  The appropriate connection
is actually
$$\eqalign{\nabla^{(1,0)} & =\delta^{(1,0)}-{1\over 4}\left(\d J\omega^{-1}
\right)^{\un i \un j}{D\over D x^{\un i}}{D\over D x^{\un j}}. \cr
        \nabla^{(0,1)} & =\delta^{(0,1)}. \cr}\eqn\corcon$$
Indeed, the commutator  $[\nabla^{(0,1)},D_{\ov i}]$ is a linear combination
of the $D_{\ov j}$, as one can see by using \oncon.  The commutator
$[\nabla^{(1,0)},D_{\ov i}]$ vanishes.  To see this, one uses, in addition
to \oncon, the defining relation
$$[D_{\ov i},D_{\un j}] = - i\omega_{\ov i \un j} \eqn\orvon$$
of the prequantum line bundle ${\cal L}$, and the relation
$$ (\d J\omega^{-1})^{ij}=(\d J\omega^{-1})^{ji}, \eqn\norvon$$
which follows from differentiating \mcon.

So $\nabla$ descends to a connection on ${\cal H}^Q$.  Now let us
compute the curvature of $\nabla$.  The $(0,2)$ part of the curvature
vanishes trivially, since $\nabla^{(0,1)}=\delta^{(0,1)}$.  The
$(2,0)$ part of the curvature can be seen to vanish using
$[D_{\un i},D_{\un j}]=0$ and also \oncon.  Let us work out in detail
the $(1,1)$ part of the curvature.  This is
$$\left[\nabla^{(0,1)},\nabla^{(1,0)}\right]
=\left[\delta^{(0,1)},-{1\over 4}\left(\d J\omega^{-1}\right)^{\un i \un j}
D_{\un i }D_{\un j} \right]. \eqn\ucucc$$
The only $J$ dependence that $\delta^{(0,1)}$ can act on is in the projection
operators implicit in the definition of the indices $\un i, \un j$.  So
we make those projection operators explicit:
$$\left(\d J\omega^{-1}\right)^{\un i \un j}
D_{\un i }D_{\un j}=\left(\d J\omega^{-1}\right)^{i j}{1\over 2}
(\delta^{i'}{}_i-iJ^{i'}{}_i)
{1\over 2}(\delta^{j'}{}_j-iJ^{j'}{}_j)
D_{i' }D_{j'}. \eqn\cncnn$$
Inserting this in \ucucc, we get
$$\left[\nabla^{(0,1)},\nabla^{(1,0)}\right]
= {i\over 8}\left(\d J\omega^{-1}\right)^{\underline i\underline j}
\delta^{(0,1)}J^{i'}{}_i
\left(D_{i'}D_j+D_jD_{i'}\right).\eqn\hdh$$

To proceed further, we restrict $\nabla$ and its curvature form
to ${\cal H}^Q$.
According
to \concon, the only non-zero components of $\delta^{(0,1)}J^{i'}{}_j$
are $\d J^{\ov i'}{}_{\un j}$.  The right hand side of \hdh\ can
therefore be simplified using the fact that
$D_{\ov i'}$ annihilates sections of ${\cal H}^Q$
and using $[D_{\ov i'},D_j]=-i\omega_{\ov i' j}$.  One gets finally
$$\left[\nabla^{(0,1)},\nabla^{(1,0)}\right]=-{1\over 8}\d J^{\un i}\,
{}_{\ov k}
\d J^{\ov k}{}_{\un i }. \eqn\mormor$$

Thus, the curvature is not zero, even when restricted to ${\cal H}^Q$.
But it is a $c$-number, that is, it depends
only on $J$ and not on the variables $x^i$ that are being quantized.
It is possible to eliminate this central curvature by adding
to $\nabla$ a one-form that depends on $J$ only or -- to formulate
this more invariantly --
by tensoring ${\cal H}^Q$ by the pullback of a line bundle on ${\cal M}$
endowed with a connection whose curvature is minus that of $\nabla$.

The fact that the curvature of $\nabla$ is a $c$-number means that
parallel transport by $\nabla$ is unique up to a scalar factor
(which moreover is of modulus 1 since the curvature is real or more
fundamentally since $\nabla$ is unitary).  So up to this factor
one can identify the various ${\cal H}_J$'s, and regard them as different
realizations of ``the'' quantum Hilbert space ${\cal H}$.

\section{Application To Calabi-Yau Manifolds}

The symplectic manifold that we want to quantize is the linear space
${\cal W}=H^3(X,\IR)$, $X$ being a Calabi-Yau threefold.  On ${\cal W}$
there is a natural symplectic form,
$$\omega(\alpha,\beta)=\int_X\alpha\wedge \beta. \eqn\opo$$
Every complex structure on $X$ determines a complex structure on ${\cal W}$,
which can be used to quantize ${\cal W}$.  So we get a family of
quantizations of ${\cal W}$, parametrized by the Teichmuller space ${\cal T}$
of complex structures on $X$ up to isotopy.
We will see that the natural connection on the family of quantum Hilbert
spaces over ${\cal T}$ is the anomaly equation of Bershadsky et. al.

\REF\xenia{P. Candelas and X. C. de la Ossa,  ``Moduli Space Of
Calabi-Yau Manifolds,'' Nucl. Phys. {\bf B355} (1991) 455.}
First, we recall some facts about variation of Hodge structures on
$X$.  (A convenient reference is [\xenia].)
The usual complex structure on ${\cal T}$ is the one in which
the $(1,0)$ part of a variation $\delta J$ of the complex structure of $X$
is $\delta J^{\un i}{}_{\ov j}$ -- or differently put, in which the
holomorphic tangent space to ${\cal T}$ is the $\overline\partial
$ cohomology group $H^1(X,T^{1,0}X)$.
Let $\Omega$ be a holomorphic three-form on $X$ that varies holomorphically
in $t$.  A basis of the complexification of ${\cal W}$ is given by
$$\eqalign{
 V_0 & = \Omega \cr
 V_a & ={\partial \Omega\over \partial t^a} \cr
 \overline V_a & = {\partial \overline\Omega\over\partial \ov t^a} \cr
 \ov V_0 & = \ov \Omega. \cr}\eqn\umigumi$$
In practice, we will be working near some
base-point $t\in {\cal T}$.  One can normalize $\Omega$
so that, at $t$, the $V_a$ are of type $(2,1)$ (and hence the $\ov V_b$ of
type $(1,2)$).  If this is done, then $\omega$ becomes block diagonal,
the non-zero matrix elements being $\omega(V_0,\ov V_0)$ and
$\omega(V_a,\ov V_b)$.  The latter are related to the
natural metric $g_{\un a\ov b}$ on $H^{2,1}(X)$:
$$\omega(V_a,\ov V_b)=\int_X{\partial \Omega\over\partial t^a}
\wedge {\partial \ov \Omega\over \partial \ov t^b} =ig_{\un a \ov b}.
\eqn\hohum$$
We will use on ${\cal T}$ a similar notation to that which we used
on ${\cal M}$ -- the $(1,0)$ and $(0,1)$ projections of a vector
$v^a$ will be written as $v^{\un a}$, $v^{\ov a}$.

The metric $g_{\un a\ov b}$ in \hohum\ is closely related to the
Zamolodchikov metric $G_{\un a \ov b}$ on the space $H^1(X,T^{1,0}X)$ of
physical states.  In fact, there is a natural map from $H^1(X,T^{1,0}X)$
to $H^{2,1}(X)$ by contracting with a holomorphic three-form; $G$ and $g$
are related by this map, so $g_{\un a\ov b}=G_{\un a\ov b}e^{-K}$,
where $e^{-K}$ is the natural metric on the space of holomorphic
three-forms.

The Yukawa couplings are
$$C_{\un a\,\un b\,\un c}
=-\int_X \Omega\wedge {\partial^3\Omega\over\partial t^a\partial
t^b\partial t^c}=\int_X{\partial \Omega\over\partial t^a}\wedge
{\partial^2 \Omega\over\partial t^b\partial t^c}\eqn\hurryy$$
(and other projections of $C$ vanish).
The second derivative $\partial^2\Omega/\partial t^a\partial t^b$ is
a linear combination of forms of type $(3,0)$, $(2,1)$, and $(1,2)$.
A non-vanishing contribution on the right hand side of \hurryy\ comes
only from the $(1,2)$ part,
which is necessarily a linear combination of the $(1,2)$ forms
$\partial \ov \Omega
/\partial \ov t^b$.  Comparing coefficients, we find
$${\partial^2\Omega\over\partial t^a\partial t^b}=-iC_{\un a\,\un b\,\un
c}g^{\un c\ov c}
{\partial \ov \Omega\over\partial \ov t^c}\;\;\;\;{\rm mod}\;\;
\d(\dots)\oplus H^{(2,1)}\oplus H^{(3,0)}.\eqn\jungol$$

We still need a few more formulas.  First of all, since $\Omega$
is of type $(3,0)$ and $J$ acts on an index of type $(1,0)$ or $(0,1)$
as multiplication by $i$ or $-i$, one has
$$J^{i'}{}_iJ^{j'}{}_jJ^{k'}{}_k\Omega_{i'j'k'}=-i\Omega_{ijk}. \eqn\ungol$$
Differentiating this and using the fact that $\partial_a\Omega$ is of type
$(2,1)$,  we get
$$ {\partial J^{i'}{}_i\over \partial t^a}\Omega_{i'jk}
+{\rm cyclic\;\;permutations\;\; of }\,\,ijk
=2i{\partial \Omega_{ijk}\over\partial t^a}.
         \eqn\nurko$$
Similarly, using the fact that that $\partial_b\Omega$ is of type $(2,1)$,
we have
$$J^{i'}{}_iJ^{j'}{}_jJ^{k'}{}_k\partial_b\Omega_{i'j'k'}=i\partial_b
\Omega_{ijk}.  \eqn\burko$$
Differentiating this, we get
$$ {\partial J^{i'}{}_i\over \partial t^a}{\partial\Omega_{i'jk}\over
\partial t^b}
+{\rm cyclic\;\;permutations\;\;of}\,\,\,ijk
=2i{\partial^2\Omega_{ijk}\over\partial t^a
\partial t^b}\;\;\;{\rm mod}\,\,\,\,H^{2,1}\oplus H^{3,0}. \eqn\murko$$
Combining this with \jungol, we have
$${\partial J^{i'}{}_i\over \partial t^a}{\partial\Omega_{i'jk}\over
\partial t^b}+{\rm cyclic\;\; permutations}
=2 C_{\un a\,
\un b\,\un c}g^{\un c\bar c}{\partial \ov \Omega\over\partial\ov t^c}
{\rm mod}\,\,\,\d(\dots)\oplus H^{2,1}\oplus H^{3,0}. \eqn\surko$$
The complex conjugate of this formula asserts that
$${\partial J^{i'}{}_i\over \partial \ov t^a}{\partial\ov\Omega_{i'jk}\over
\partial \ov t^b}+{\rm cyclic\;\; permutations}
=2\ov  C_{\ov a\ov b\ov c}g^{\un c\bar c}{\partial \Omega\over\partial t^c}
{\rm mod}\,\,\,\d(\dots)\oplus H^{1,2}\oplus H^{0,3}. \eqn\ssurko$$

\section{Complex Structure Of ${\cal W}$}

We now must study the complex structure of ${\cal W}$.
An element $\Theta$ of ${\cal W}=H^3(X,\IR)$ has an expansion
$$\Theta=\lambda^{-1}\Omega+\sum_au^a{\partial \Omega\over\partial t^a}
+\sum_a\ov u^a{\partial\ov \Omega\over\partial\ov t^a}+\ov\lambda^{-1}\ov
\Omega \eqn\cnson$$
in the basis \umigumi, with complex coefficients $\lambda$, $u^a$.
The object $\lambda$ appearing here will turn
out to be the string coupling constant.

We now have to make explicit the complex structure of ${\cal W}$.
${\cal W}$ has no natural complex structure, but every choice of
complex structure $J$ on $X$ enables one to pick a complex structure,
which I will call ${\cal J}$, on ${\cal W}=H^3(X,\IR)$.
We choose ${\cal J}$ to be the complex
structure on $H^3(X,\IR)$ in which
$H^{3,0}$ and $H^{2,1}$ are considered to be of type $(1,0)$.
Hence, ${\cal J}$
should be represented by an operator on differential forms that multiplies
$(3,0)$ and $(2,1)$ forms by $i$, and multiplies $(0,3)$ and $(1,2)$ forms
by $-i$.  Such an operator is
$$({\cal J}\Theta)_{ijk}={1\over 2}J^{i'}{}_iJ^{j'}{}_jJ^{k'}{}_k
\Theta_{i'j'k'}+{1\over 2}\left(J^{i'}{}_i\Theta_{i'jk}+{\rm cyclic
\;\;permutations\;\; of}\,\,\,ijk\right). \eqn\xonx$$

We now need to compute the $\ov t$ dependence of ${\cal  J}$.
As in \S1, we will use the notation $t'$ for $\ov t$.
The computation of $t'$ dependence is straightforward:
$$\left({\partial {\cal J}\over\partial t'{}^a}\Theta\right)_{ijk}
={1\over 2}\left({\partial J^{i'}{}_i\over\partial t'{}^a}J^{j'}{}_j
J^{k'}{}_k\Theta_{i'j'k'}+{\partial J^{i'}{}_i\over \partial t'{}^a}
\Theta_{i'jk}\right)+{\rm cyclic\;\;permutations\;\;of}\;\; ijk.
 \eqn\mcnon$$
In particular, if $\Theta$ is of type $(1,2)$, this implies
$$\left({\partial {\cal J}\over\partial t'{}^a}\Theta\right)_{ijk}
 = \left({\partial {J}^{i'}{}_i\over\partial t'{}^a}\Theta_{i'jk}
+{\rm cyclic\;\;permutations}\right)\,\,\,\,{\rm mod}\,\,H^{1,2}\oplus H^{0,3}.
\eqn\jcncn$$
Taking $\Theta=\partial\ov\Omega/\partial t'{}^b$,
and using \ssurko, we learn that
$${\partial {\cal J}\over \partial t'{}^a}\left(
{\partial\ov \Omega\over \partial
t'{}^b}\right)=2\ov  C_{\ov a\ov b\ov c}g^{\un c\bar c}
{\partial \Omega\over\partial t^c}
\;\;\;{\rm mod}\,\,\,\d(\dots)\oplus H^{1,2}\oplus H^{0,3}.
\eqn\oodpn$$

\section{Final Evaluation}

To make completely explicit the connection \corcon\ for quantization
of ${\cal W}$, with a family of polarizations parametrized by ${\cal T}$,
we need to evaluate $(\d {\cal J}\omega^{-1})^{\un i\un j}$.  The only
non-zero matrix element (given that $\un i$ and $\un j$ are to be indices
of type $(1,0)$, corresponding to $(3,0)$ or $(2,1)$ forms) comes from
\oodpn\ together with
$$\omega^{-1}{}^{\ov b \un a}=-ig^{\ov b \un a}, \eqn\usnon$$
which is equivalent to \hohum.    Combining these pieces, we get
$$(\d {\cal J}\omega^{-1})^{\un i\un j}D_{\un i}D_{\un j}
=-2i \sum_a\d t'{}^a \ov C_{\ov a\ov b\ov c} g^{\un b\bar b}g^{\un c\bar c}
{D\over Du^b}
{D\over D u^c}.          \eqn\rusnon$$

The condition that the quantum state represented by a vector $\Psi(\lambda,
u;t')$ is independent of $t'$ can be read off from \corcon\ and
is
$$\eqalign{
\left({\partial\over\partial t'{}^a}+{i\over 2}\ov
C_{\ov a\ov b\ov c}g^{\un b\bar b}g^{\un c\bar c}
          {D\over Du^b}{D\over Du^c}\right)\Psi & = 0 \cr
                           {\partial\over \partial {\bar t}'{}^a}\Psi & = 0.}
       \eqn\longwork$$
The main point of this paper is that the first equation in \longwork\
practically coincides
with the holomorphic anomaly equation \hiko\ of Bershadsky et. al.\foot{The
second equation in \longwork\ is also true in their formalism.
They consider $t$ and $\bar t$ as independent complex variables and
consider functions that are holomorphic in $t,\bar t$.  Given that
their $t,\bar t$ correspond to our $u,t'$, holomorphicity in $\bar t$ is the
second equation in \longwork\ and holomorphicity in $t$ is \jsnsn.}
A factor of $2i$ presumably results from a difference in conventions
(for instance, as will be clear momentarily, it can be absorbed in
the definition of the string coupling constant).  Let us analyze the remaining
discrepancies.

First of all, Bershadsky et. al.
consider the partition function as a function of the
string coupling constant and also the complex structure of $X$.
In quantizing ${\cal W}$, our wave functions depend on the variables
$\lambda$ and $u$ introduced  in
\cnson.  $\lambda$ determines a holomorphic three-form, and in the $B$ model
this means that $\lambda$ should be associated with the string coupling
constant.  On the other hand, $u$ is a $(2,1)$ cohomology class.
To Bershadsky et. al., the natural variables would be the string
coupling constant $\lambda$ and an element $t$ of $H^1(X,T^{1,0}X)$.\foot{
Recall from \S1 that once a base-point is picked, the Teichmuller space
of $X$ has a natural local isomorphism with the linear space $H^1(X,T^{1,0}X)$
via special coordinates.}
In a familiar fashion, one can map $H^1(X,T^{1,0}X)$ to $H^{2,1}(X)$
by contracting with a holomorphic three-form, which in this case
should be naturally the one determined by $\lambda$.  This means that
the natural relation between $u$ and $t$ is $t=\lambda u$.
In terms of $t$, the first equation in \longwork\ would therefore be
$$\left(
{\partial\over\partial t'{}^a}+{i\over 2}
  \lambda^2\ov C_{\ov a\ov b\ov c}g^{\un b\bar b}g^{\un c\bar c}
          {D\over Dt^b}{D\over Dt^c}\right)\Psi = 0. \eqn\sosn$$
Now we see the natural appearance of the string coupling constant, as in
\hiko.

Another important point is that the connection \corcon\ for quantization
of an affine space is only projectively flat.  Therefore, in attempting
to study the $t'$ dependence of a wave function $\Psi$ using
\longwork, there would be an undetermined factor of modulus unity --
a $c$-number factor in the sense that it depends only on $t'$ and not on $t$.
Of course, this comes from the fact that $\Psi$
is really a section of the prequantum line bundle rather than a function.
Perhaps a trivialization of this line bundle is implicit
in [\vafa].

In the definition of
the partition function $Z=\exp({1\over 2}\sum_{g=0}^\infty\lambda^{2g-2}F_g)$,
the genus one term $F_1$ has the following characteristics: (i) the power
of $\lambda$ multiplying it vanishes; (ii)
because of zero mode contributions (analogous to those in Ray-Singer
analytic torsion) it is not most naturally interpreted as a ``function''
but as a section of a certain line bundle.
A better understanding of $F_1$ might clarify the role of the
prequantum line bundle.

Despite unresolved questions, the resemblance of the holomorphic
anomaly equation to the equation \sosn\ of quantum background independence
is so close that it is hard to believe that it is a coincidence.
This relation seems likely to repay further study.

\section{A Speculation}

For instance, it is amusing to speculate along the following lines.
Perhaps the phase space ${\cal W}$ that has appeared in this discussion
should be interpreted as part of the usual classical
phase space of a system, and the construction of the auxiliary Hilbert
space ${\cal H}$ should be regarded as part of the process of quantization.
Then what is unusual is that -- as opposed to the usual situation
in which the choice of a vector in ${\cal H}$ is a choice of initial
conditions -- the physical system here determines a distinguished vector
in ${\cal H}$, namely the partition function.

\REF\hawking{S. Hawking, ``The Cosmological Constant Is Probably Zero,''
Phys. Lett. {\bf 13B} (1984) 403.}
Before identifying this as a cosmologist's dream, in which the
initial conditions of the universe are uniquely determined by fundamental
theory, we should pause to note that in physical applications of string
theory, background independence is realized ``normally'' and one
probably does not want to abandon that since the realization of background
independence in general relativity is ``normal.''
``Quantum'' background independence
as we have encountered it in this paper apparently depends on
having a non-trivial cohomology of the $b_0,\ov b_0$ operators (which
leads to considerations of $t-\ov t$ fusion and the holomorphic anomaly);
this is absent in the usual critical string theories.  Though one
probably would not want ``quantum'' background independence
for transverse gravitons, perhaps there is a modification of the usual
string theories in which the BRST cohomology is such that ``quantum''
background independence holds for the conformal factor in the space-time
metric.  Then -- blindly imitating what we have found above -- quantum
background independence might dictate that the dependence of the wave
function on the conformal factor should be given by the partition function.
It has been argued [\hawking] that under such conditions the
cosmological constant would vanish, since under some hypotheses
the partition function of the universe diverges at zero cosmological constant.

\REF\witz{E. Witten and B. Zwiebach, ``Algebraic Structures And Differential
Geometry In 2-D String Theory,'' Nucl. Phys. {\bf B377} (1992) 55.}
\REF\verr{E. Verlinde,  ``The Master Equation of 2-D String Theory,''
Nucl. Phys. {\bf B381} (1992) 141.}
\REF\zwiebach{B. Zwiebach, ``Closed String Field Theory: Quantum Action
And The B-V Master Equation,'' Nucl. Phys. {\bf B 390} (1993) 33.}
In any event, though ``ordinary''
background independence appears to suffice (apart from such exotic
speculations)
for the usual physical
applications of string theory, familiarity with ``quantum'' background
independence may be useful in trying to go off-shell.  I am reminded
of the BV formalism of quantization, which enters on-shell only
in exotic string theories in $D\leq 2$ [\witz,\verr],
but seems to be very valuable
in formulating critical string theories off-shell, even at the classical
level [\zwiebach].

\ack{I would like to thank P. Aspinwall, M. Bershadsky, H. Ooguri,
D. Morrison, and C. Vafa for discussions and helpful suggestions.}
\refout
\end